# Defying Diamagnetism: Anomalous Redshift and Robust Chiral Spin Mixing in 2D Perovskites under High Magnetic Fields up to 50 T

*Dae Young Park[1], Sein Park[2], Yongmin Kim[2*], Nojoon Myoung[3], Zhuo Yang[4], Yoshimitsu Kohama[4], Yasuhiro H. Matsuda[4],*

[1]Department of Science Education, Jeonju University, Jeonju 55069, Korea

[2]Department of Physics, Dankook University, Cheonan 31116, Korea

[3]Department of Physics Education, Chosun University, Gwangju 61452, Korea

[4]Institute for Solid State Physics, The University of Tokyo, Kashiwa 277-8581, Japan

**Corresponding Author**

*E-mail: yongmin@dankook.ac.kr

Phone: +82-41-550-3422    Fax: +82-41-559-7859

ABSTRACT

We report circularly polarized magneto-photoluminescence (MPL) measurements on the enantiomeric pair of two-dimensional chiral perovskites *R*- and *S*-$(C_4H_9NH_3)_2PbI_4$ in pulsed magnetic fields up to 50 T at 2.0 K. The *R*-form exhibits strong preferential emission into the right circular (RC) polarized light (R-CPL) ($I_{RC}/I_{LC} \approx 6.0{:}1$), with the *S*-form showing the opposite sense, robust against the applied field. Three key observations are reported: (i) an anomalous redshift of all exciton transition energies with increasing field - contrasting with the normal diamagnetic blueshift of achiral perovskites - with a non-monotonic initial blueshift to ~7.6 T in the dominant RC polarization of the *R*-form; (ii) complete invariance of PL peak positions upon reversal of the field direction, consistent with Rashba-induced mixing of the bright exciton with a nearby $m_x = 0$ dark state; and (iii) field-induced increase in PL intensity and narrowing of the linewidth, attributed to suppression of disorder scattering as the cyclotron orbit contracts. These results are interpreted in the framework of Rashba spin-splitting and polaron formation in the chiral two-dimensional perovskite.

---

Two-dimensional (2D) organic–inorganic hybrid perovskites (2D-HOIPs) of the form $(RNH_3)_2PbX_4$ (R is an alkyl and X = I, Br, Cl) form natural quantum-well heterostructures in which the $[PbX_6]^{4-}$ inorganic sheets are sandwiched between insulating organic cation bilayers. Strong dielectric confinement amplifies the exciton binding energy to several hundred meV,[1,2,3,4] and the heavy-metal Pb 6*p* character of the conduction-band minimum provides exceptionally large spin-orbit coupling (SOC).[5,6,7] These properties make 2D-HOIPs versatile platforms for exciton physics and for exploring SOC-related phenomena including the Rashba effect[8,9] and Fröhlich large-polaron formation.[10]

The introduction of enantiopure chiral ammonium cations as the organic spacer layer breaks inversion symmetry of the crystal and enables structural chirality transfer to the inorganic $[PbX_6]^{4-}$ sublattice. Kim et al.[11] demonstrated, using DFT simulations of chiral methylbenzylammonium (MBA) ligands on $CsPbBr_3$ nanocrystal surfaces, that the chiral organic cations induce a centro-asymmetric distortion of the Pb-Br-Pb bond angles penetrating up to five unit cells into the inorganic lattice, which is proposed as the structural origin of the chiroptical response.[12,13,14] In 2D chiral perovskites based on *R*- and *S*-MBA lead iodide, this structural symmetry breaking produces both circular dichroism (CD) and circularly polarized photoluminescence (CPPL), with the chiroptical activity amplified when a higher degree of inorganic octahedral tilting (i.e., distortion of the Pb–I–Pb bond angle) promotes stronger asymmetric hydrogen bonding between the chiral cation and the inorganic framework.[15,16,17] The potential of this platform for device applications has been demonstrated by circular polarized light (CPL) photodetectors based on quasi-2D chiral perovskite films using naphthylethylammonium (NEA) cations, achieving a dissymmetry factor of 0.15 and a responsivity of 15.7 A $W^{-1}$.[18]

Magneto-optical spectroscopy is a powerful probe of exciton fine structure, effective mass, and spin textures in semiconductors.[19] In 3D methylammonium lead trihalide perovskites ($MAPbX_3$), high-field MPL studies have revealed a halide-dependent interplay between Rashba band inversion and polaron renormalization that leads to unconventional, non-diamagnetic energy shifts - including quadratic redshifts ($MAPbCl_3$) and power-law field dependences ($MAPbI_3$) distinct from the normal diamagnetic blueshift.[19,20,21] For 2D chiral systems, the additional constraints of chirality-imposed inversion symmetry breaking are expected to modify the exciton response in qualitatively new ways; recent theoretical work has shown that CD and CPL in chiral 2D perovskites emerge from spin-splitting via cross-coupling of Rashba-like and chiral spin-texture components.[22,23,24,25,26,27,28,29]

Here, we present a systematic MPL study of the enantiomeric pair $R$- and $S$-$(C_4H_9NH_3)_2PbI_4$ using circularly polarized detection in pulsed fields up to 50 T. We report: (i) strong preferential CPL emission with opposite helicity for the two enantiomers, (ii) anomalous monotonic redshift of both left circular (LC)- and right circular (RC)-polarized transitions in the $S$-form, (iii) non-monotonic (initial blueshift followed by redshift) behavior of the RC-polarized transition in the $R$-form, (iv) field-reversal invariance of PL energies, consistent with Rashba-induced mixing of the bright exciton with a nearby $m_x = 0$ dark state, and (v) field-induced narrowing of the linewidth in all cases. We discuss these observations in terms of Rashba spin splitting, polaron formation, and the symmetry constraints of the chiral lattice.

Figure 1 shows the circularly polarized PL spectra of the $R$- and $S$-form crystals at 2.0 K for zero field ($B = 0$ T) and at $B = 50$ T, measured with RC and LC polarizers. For this experiments, the magnetic fields were fixed to $+\hat{z}$ (↑) direction normal to the sample plane (Faraday configuration). Circular polarizer was inserted between the sample and the optical fiber. The detailed experimental setup can be found elsewhere.[19] The *R*-form sample [Fig. 1(a)] shows a dominant PL peak near 2.314 eV in the RC polarization and a strongly suppressed peak near 2.336 eV in the LC polarization at zero field. The integrated intensity ratio $I_{RC}/I_{LC}$ is approximately 6.0:1 at $B = 0$ T and remains similarly large (~9:1) at 50 T, demonstrating that the preferential emission into the RC polarization is robust against the applied field and is therefore a consequence of the intrinsic chiral crystal symmetry. The substantial energetic separation between the RC and LC peaks in the *R*-form (~21 meV at zero field) reflects the large CPL splitting induced by the chiral crystal field.

The *S*-form sample [Fig. 1(b)] shows the mirror-image behavior: dominant emission in the LC polarization at ~2.326 eV and suppressed emission in the RC polarization at ~2.329 eV. The LC/RC intensity ratio is approximately 2.4:1 at zero field, increasing to ~3.1:1 at 50 T. The energetic splitting between the two polarization components in the *S*-form (~3 meV) is considerably smaller than in the *R*-form, indicating that the two enantiomers, while exhibiting opposite helicities, do not necessarily display identical CPL splitting magnitudes - a feature that may reflect differences in the degree of crystal perfection or octahedral tilting angle between the two samples.

Under the applied field of 50 T, the PL peaks in both enantiomers shift to lower energy relative to their zero-field positions (see also Figs. 3 and 4), while the dominant polarization retains its preferential intensity. The observation that the RC peak remains dominant in the *R*-form at 50

T indicates that the chiral selection rules are not quenched by the external magnetic field at this field strength.

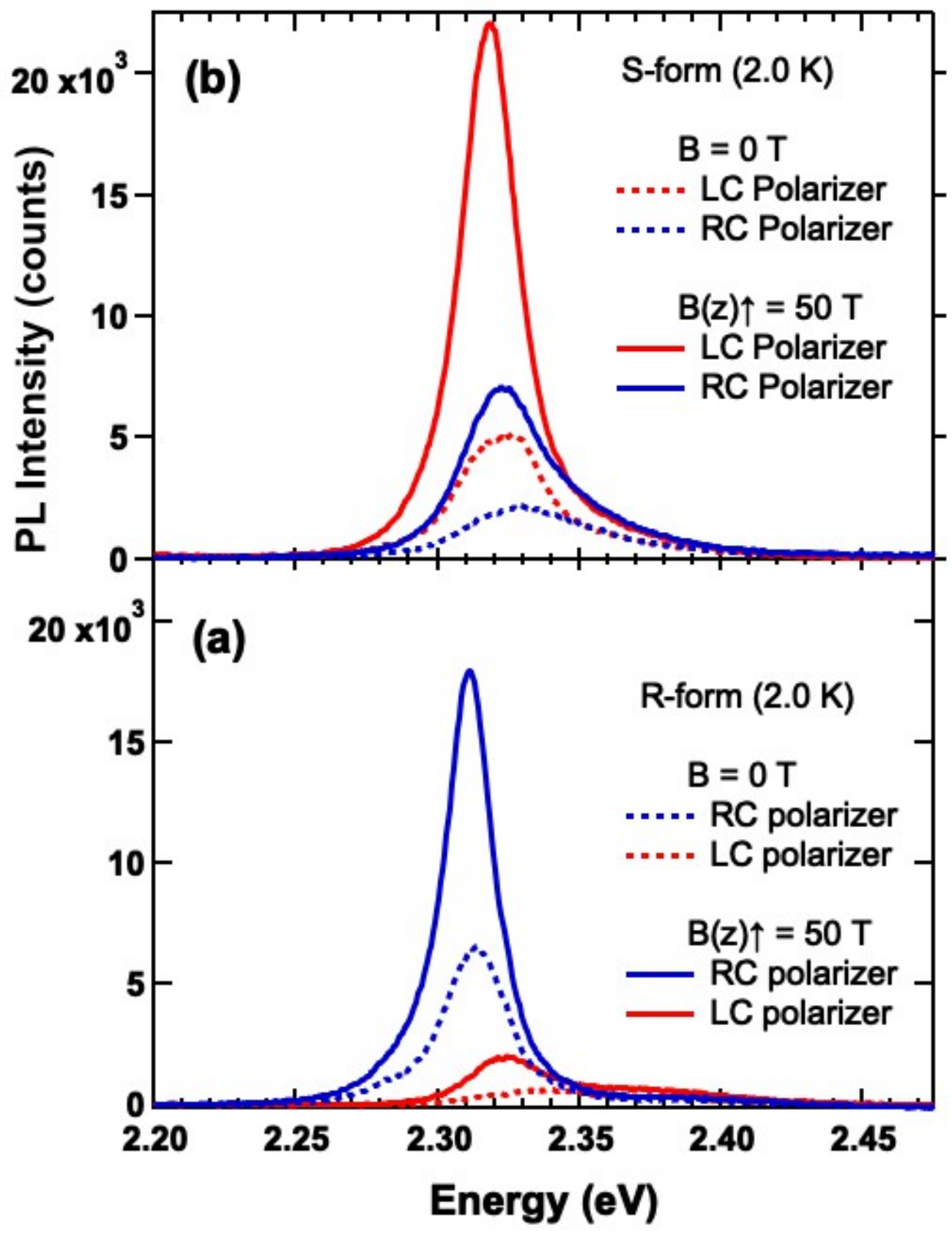

**Figure 1.** Circularly polarized PL spectra of (a) $R$-form and (b) $S$-form $(C_4H_9NH_3)_2PbI_4$ at $T = 2.0$ K for $B = 0$ T (dashed lines) and $B = 50$ T (solid lines), measured with RC (blue) and LC (red) polarizers. The $R$-form shows dominant RC emission (~6.0:1 intensity ratio) and the $S$-form shows dominant LC emission (~2.4:1 intensity ratio), with opposite helicity in each case at zero field.

Figure 2 shows the PL spectra of the $R$- and $S$-form samples at $B = 0$ T and at $B = 50$ T for both field directions ($\boldsymbol{B}(+\hat{z})$↑ and $\boldsymbol{B}(+\hat{z})$↓), measured with the preferred circular polarizer for each sample (RC for $R$-form, LC for $S$-form). In both enantiomers, the PL peak position and shape are identical for the two field directions at $B = 50$ T, within our spectral resolution.

This field-direction invariance is a key observation. In conventional Zeeman splitting, reversing the field direction exchanges the energies of the $\sigma^+$and $\sigma^-$ exciton states, producing a symmetric spectral shift. The absence of such a shift here implies that the relevant exciton transition is not a pure Zeeman doublet with quantum number $m_s = \pm 1$, but rather a state whose energy is set by an internal interaction that does not change sign upon field reversal.

A closely related field-direction invariance has been identified for a specific exciton transition in single $CsPbBr_3$ nanocrystals by Isarov et al.[9], wherein the calculated circularly polarized PL spectra from the Rashba-Zeeman exciton model show that the low-energy exciton transition ($E_{Low}$) has a very weak dependence on magnetic field strength. The authors attribute this to mixing of the $E_{Low}$ bright exciton ($J_x = 1, m_x = \pm 1$) with a nearby dark state of $J_x = 1, m_x = 0$ spaced ~1 meV away, where $J_x$ denotes the exciton total angular momentum and $m_x$ its projection along the Rashba symmetry axis $x$. Because the $m_x = 0$ state carries no net angular momentum along the field axis; consistent with this, Isarov et al.[9] found its energy to depend only weakly on the field magnitude (up to 8 T). We propose that, by the same symmetry argument, such a state's energy should also be insensitive to the sign (direction) of the applied field. The mixing of the $E_{Low}$ bright exciton with this $m_x = 0$dark state, driven by the Rashba-induced inversion symmetry breaking, is the mechanism we propose could analogously pin the $E_{Low}$ transition energy and its circular polarization character against reversal of the field direction - an extrapolation beyond what Isarov et al.[9] themselves reported, since their study was confined to field magnitudes up to 8 T and did not examine field-direction reversal. In our chiral $(C_4H_9NH_3)_2PbI_4$ system, the structural inversion symmetry breaking induced by the chiral ammonium cation introduces Rashba-type spin-orbit coupling into the $[PbI_6]^{4-}$ sublattice, promoting analogous mixing between the bright exciton and the nearby $m_x = 0$ state. Specifically, as demonstrated by Kim et al.,[11] the chiral ammonium cations induce a centro-asymmetric distortion of the Pb-I-Pb bond angles in the inorganic sublattice; this structural distortion constitutes the local inversion symmetry breaking that activates the Rashba-type coupling responsible for mixing the $m_x = \pm 1$ bright exciton with the $m_x = 0$ dark state - providing a structural basis for the field-direction invariance observed here, though we note that this mechanism, initially identified in 0D nanocrystals by Isarov et al.,[9] has not yet been theoretically confirmed for 2D chiral systems. This mixing, enhanced by the large chiral splitting (~21 meV in the *R*-form), provides a natural explanation for the observed insensitivity of the PL peak position and circular polarization to the sign of the applied field throughout the 0 ~ 50 T range.

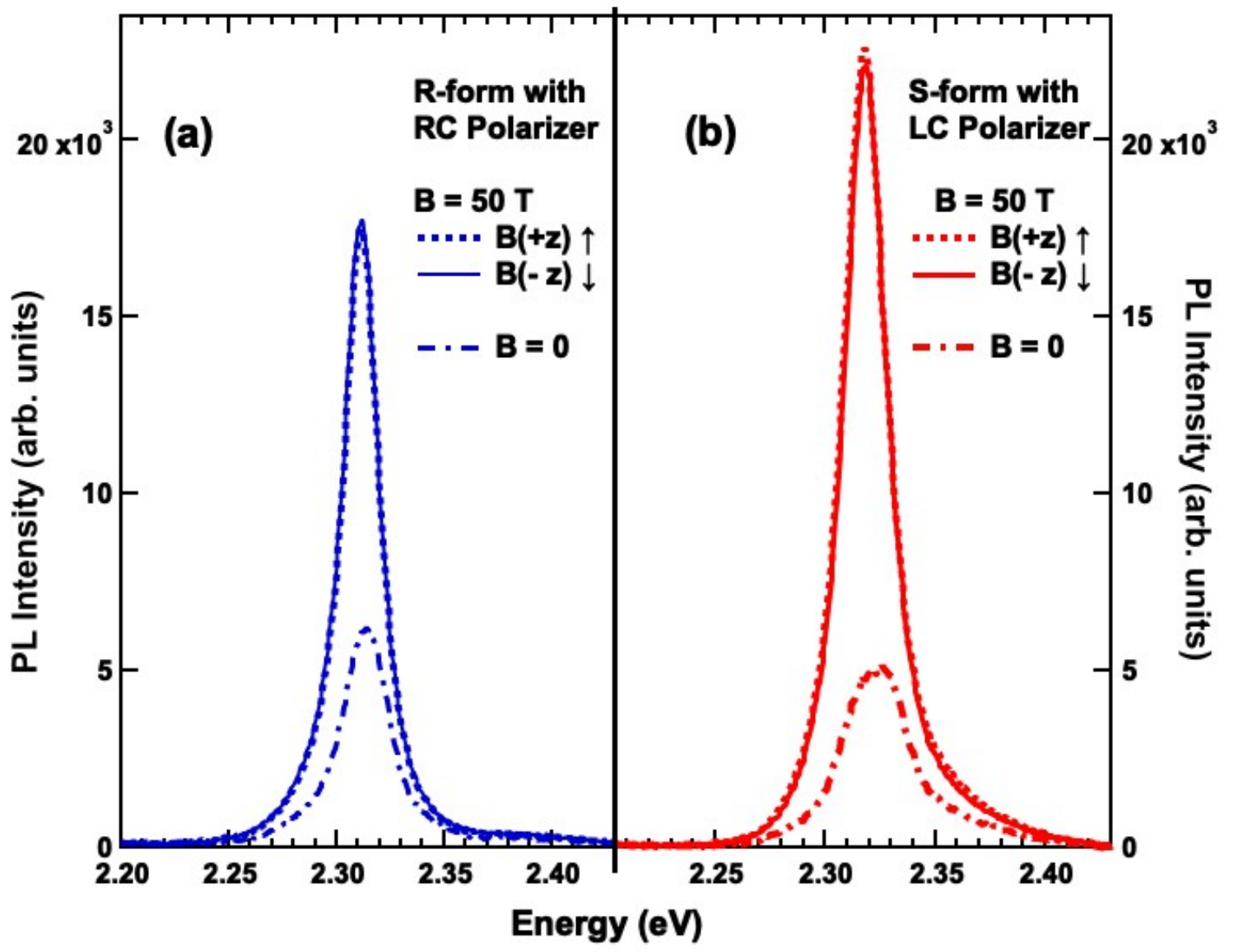

**Figure 2.** Circularly polarized PL spectra of (a) *R*-form (RC polarizer) and (b) *S*-form (LC polarizer) at $B = 0$ T (dash-dot lines) and $B = 50$ T for $\boldsymbol{B}(+\hat{z})\uparrow$ (dotted lines) and $\boldsymbol{B}(-\hat{z})\downarrow$ (solid

lines). The PL peak positions are identical for both field directions at 50 T in both samples, demonstrating magnetic field-direction invariance.

To understand the anomalous redshifts observed in Figs. 3(a) and 4(a), it is instructive to consider the polaron framework described by Shin et al.[19] for 3D $MAPbX_3$. In that work, the key criterion governing the field-dependent behavior is the comparison between the cyclotron frequency $\omega_c$ and the polaron frequency $\omega_p$. At low applied fields, $\omega_c \ll \omega_p$: the carrier remains bound to the polar lattice ions as a polaron, and the SOC-induced Rashba splitting - which is set by the internal lattice potential and is independent of the sign of the external field - dominates over the Zeeman splitting. As the field increases, the lowest Landau level cyclotron radius $r_c = \ell_B$ (where $\ell_B = \sqrt{\hbar/eB}$ is the magnetic length) shrinks and approaches the polaron radius $r_p$. When $\omega_c$ exceeds $\omega_p$, i.e., when $\ell_B \lesssim r_p$, the polaron binding softens: the carrier decouples from the polar lattice ions and the polaron-Rashba contribution to the effective mass is reduced. It is generally known that the full exciton transition-energy shift is described by

$$\Delta E = \pm\frac{1}{2}\Delta g \mu_B B + c_0 B^2 \qquad (1)$$

where the first term is the Zeeman energy and the second term is the diamagnetic shift with coefficient $c_0 = e^2 a_B^{*2}/(8\mu)$, which is proportional to the exciton Bohr radius squared ($a_B^{*2}$) and inversely proportional to the reduced effective mass $\mu$. Critically, the sign of $c_0$ depends on the sign of $\mu$: when the Rashba-polaron coupling renders $\mu$ negative[21] (as in the $MAPbCl_3$ peaks 1 and 2 at low field[19]), $c_0$ is negative and the $c_0 B^2$ term produces a redshift that grows as $B^2$. Once $\omega_c$ exceeds $\omega_p$ and the polaron softens, the effective mass $\mu$ recovers to a positive value, flipping the sign of $c_0$ from negative to positive. As a result, the $c_0B^2$ term, which scales as $B^2$, now produces a blueshift that grows rapidly with field. This is precisely the behavior observed for peak 2 of $MAPbCl_3$ in Shin et al.[19], where $c_0$ changes sign from −1.08 to +0.72 μeV $T^{-2}$ as the field crosses ~20 T. In our 2D chiral perovskite, the analogous polaron softening crossover field may lie above 50 T, consistent with the observation that both polarizations exhibit a monotonic redshift throughout the measured field range without the recovery to blueshift seen in $MAPbCl_3$.

Figure 3 shows the magnetic field dependence of (a) the peak transition energy, (b) the integrated PL intensity, and (c) the full-width at half-maximum (FWHM) for the $R$-form sample, measured separately with RC and LC polarizers. The peak transition energy [Fig. 3(a)] shows qualitatively different field dependences for the two polarization components. The RC-polarized peak, which lies near 2.314 eV at zero field, initially undergoes a small blueshift, reaching a maximum near $B \approx 7.6$ T with a shift of approximately +0.8 meV, before turning over and shifting monotonically to lower energy with increasing field, reaching ~2.3116 eV at 50 T (a total shift of −2.6 meV from zero field). In contrast, the LC-polarized peak, lying near 2.332 eV, decreases monotonically from zero field and reaches ~2.323 eV at 50 T, a total redshift of ~ −9.5 meV. Both transitions thus exhibit net redshift at high fields, with the LC polarization showing a substantially larger total shift.

The non-monotonic field dependence of the RC-polarized peak in the $R$-form - an initial blueshift of +0.8 meV up to ~7.6 T followed by a sustained redshift reaching −2.6 meV at 50 T - is a distinctive feature that has no counterpart in the LC polarized emission of the same sample nor in either polarization of the $S$-form. The turnover at ~7.6 T suggests that the RC-polarized transition in the $R$-form involves a competition between at least two field-dependent energy contributions of opposite sign that cross over near this field. The physical origin of this crossover - which may involve the interplay between the diamagnetic term, Rashba-polaron renormalization, and the chiral crystal field specific to this enantiomer and polarization component - is not yet fully understood. A detailed theoretical treatment of the magneto-optical response of chiral 2D perovskites, incorporating the combined effects of chirality, Rashba spin-orbit coupling, and polaron formation in the 2D quantum-well geometry, will be needed to account for this observation. This non-monotonic behavior of the dominant emission polarization represents one of the most intriguing aspects of the present dataset and warrants further experimental and theoretical investigation, including measurements at intermediate fields and as a function of temperature.

The integrated PL intensity [Fig. 3(b)] increases with magnetic field for both polarization components, though the RC polarization remains substantially stronger (RC/LC area ratio ~6 at zero field). Both intensities approach saturation at high fields. This enhancement of PL intensity with field is a common observation in 2D lead-halide perovskites under high magnetic fields and is generally attributed to field-induced brightening of exciton states.[30-35]

The FWHM [Fig. 3(c)] decreases monotonically with increasing field for both polarizations, from ~20.5 meV (LC) and ~12.9 meV (RC) at zero field to ~16.8 meV and ~9.6 meV respectively at 50 T. The physical origin of this linewidth narrowing in the present system is not yet established and will require further study, although analogous field-induced narrowing has been attributed to suppression of disorder-related scattering in other two-dimensional systems.

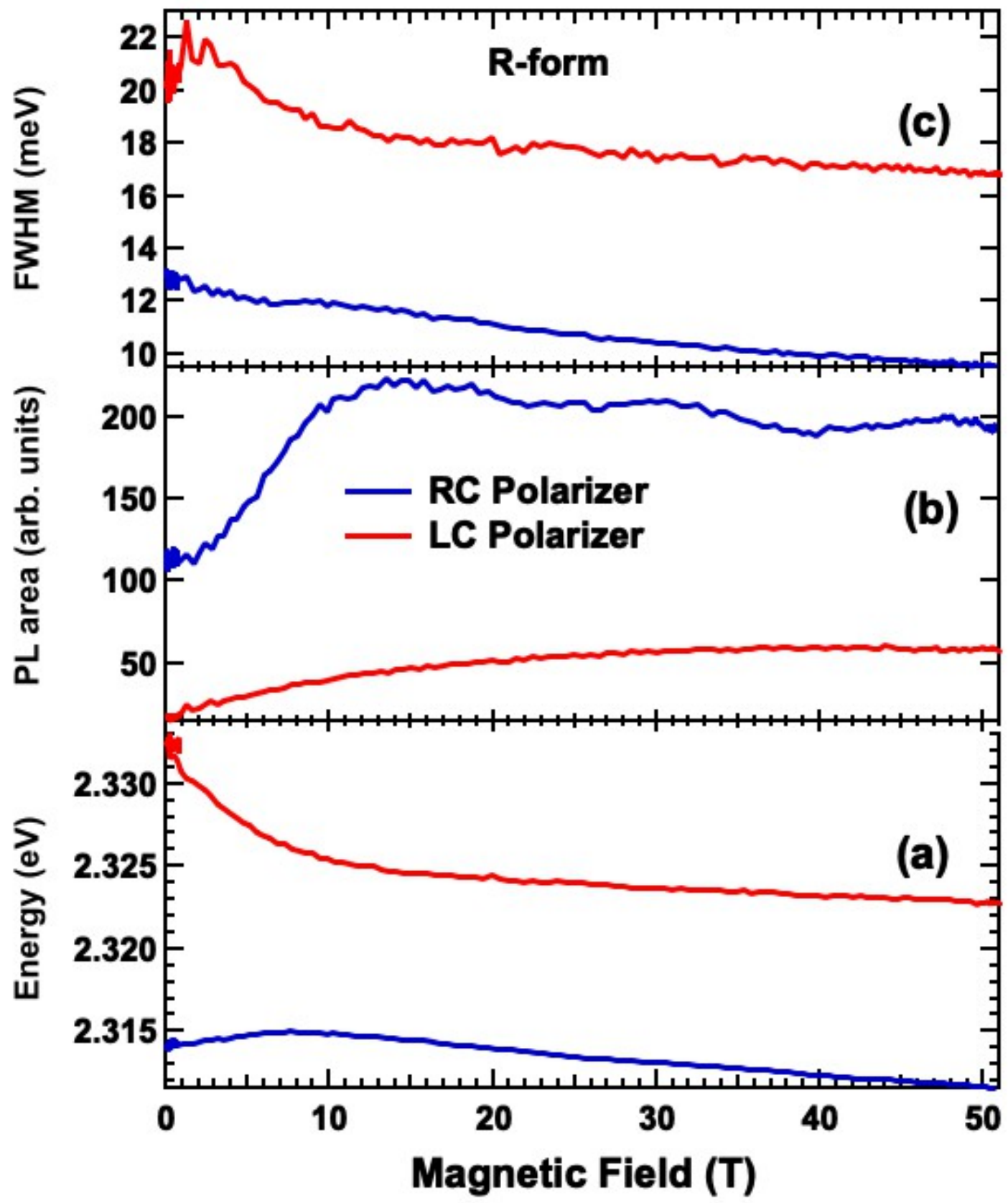

**Figure 3.** Magnetic field dependence of (a) peak transition energy, (b) integrated PL areal intensity, and (c) FWHM of the $R$-form $(C_4H_9NH_3)_2PbI_4$ sample with RC (blue) and LC (red) polarizers at $T = 2.0$ K. The RC-polarized peak shows an initial blueshift to ~7.6 T followed by a monotonic redshift, while the LC peak decreases monotonically. Both polarization components show increasing PL intensity and decreasing FWHM with field.

Figure 4 shows the corresponding data for the $S$-form sample. In contrast to the $R$-form, both the LC- and RC-polarized peaks of the $S$-form decrease monotonically with increasing field throughout the entire field range from 0 to 50 T, without the initial blueshift seen in the $R$-form RC polarization.

The LC-polarized peak (dominant polarization component of the *S*-form) starts at ~2.321 eV and shifts to ~2.318 eV at 50 T (a redshift of ~−3.0 meV). The RC-polarized peak starts at ~2.327 eV and redshifts by ~ −4.3 meV to ~2.322 eV at 50 T. Both polarizations thus show monotonic redshifts with similar magnitudes. The absence of an initial blueshift in either polarization for the *S*-form may reflect a different balance between the diamagnetic contribution and the Rashba-polaron renormalization in this enantiomer, possibly related to the smaller CPL splitting (~3 meV) compared to the *R*-form (~21 meV), which may indicate a different degree of structural distortion.

The integrated PL intensity [Fig. 4(b)] increases monotonically with field for both polarizations, with the LC polarization remaining dominant (LC/RC ratio ~2.7 at zero field, increasing to ~3.5 at 50 T). The FWHM [Fig. 4(c)] decreases monotonically for both polarizations, from ~17.9 meV (LC) and ~20.6 meV (RC) at zero field to ~13.9 meV and ~16.5 meV respectively at 50 T. These trends are qualitatively similar to the *R*-form behavior.

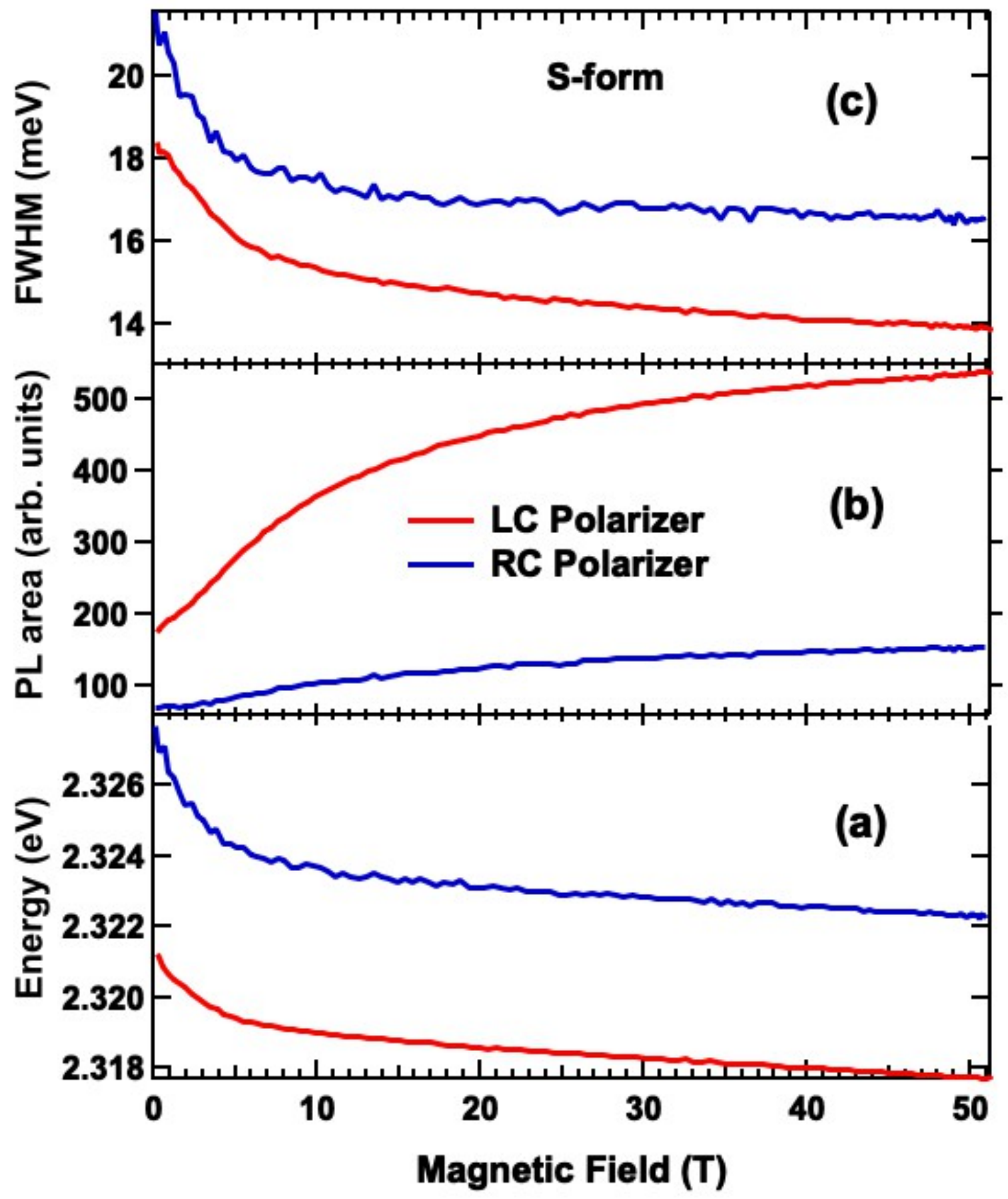

**Figure 4.** Magnetic field dependence of (a) peak transition energy, (b) integrated PL areal intensity, and (c) FWHM of the *S*-form $(C_4H_9NH_3)_2PbI_4$ sample with RC (blue) and LC (red) polarizers at $T = 2.0$ K. Both polarization polarizations show monotonic redshift with increasing field, in contrast to the non-monotonic RC- polarization behavior of the *R*-form.

The overall picture emerging from Figures 1–4 is that the chiral crystal field in *R*- and *S*$(C_4H_9NH_3)_2PbI_4$ produces strong CPL anisotropy and imposes field-direction-invariant selection rules on the exciton transitions, while the applied magnetic field drives anomalous (non-diamagnetic) energy shifts in both enantiomers. These observations can be rationalized within a framework that combines three physical effects:

**Chiral crystal field and CPL selection rules.** The asymmetric $[PbI_6]^{4-}$ octahedral distortion induced by the chiral ammonium cation produces a noncentrosymmetric polar crystal, breaking inversion symmetry and creating a large internal electric field that splits the exciton into RC- and LC-emission polarizations with energies differing by ~21 meV in the $R$-form and ~3 meV in the $S$-form. The direction of preferred emission (RC for $R$-form, LC for $S$-form) is opposite for the two enantiomers, as expected from the opposite sense of octahedral tilting. We propose that the field-direction invariance of the PL peak energies observed throughout 0 ~ 50 T can be understood as an extension of the dark-state mixing mechanism reported by Isarov et al.[9] for $CsPbBr_3$ nanocrystals (Fig. 4b of that work): the low-energy exciton transition acquires a very weak field dependence because the Rashba-induced inversion symmetry breaking mixes the bright exciton ($J_x = 1, m_x = \pm1$) with a nearby dark state ($J_x = 1, m_x = 0$). Because the $m_x = 0$ state carries no net angular momentum along the field axis, so its energy was found to depend only weakly on field magnitude up to 8 T in that work. We extend this reasoning to propose that such a state's energy should likewise be insensitive to field direction - a regime and a claim not tested by Isarov et al. themselves. Under this proposed extension, the resulting mixed state would have a transition energy pinned against field reversal. In our chiral system, the stronger inversion symmetry breaking imposed by the chiral ammonium cation would be expected to enhance this mixing, which we suggest as a plausible explanation for the field-direction invariance we observed robustly throughout the full 0 ~ 50 T range.

To make this pinning mechanism quantitative, we model the dark and bright states as two levels coupled by a Rashba-induced off-diagonal matrix element $\Delta_R$. The coupled eigenstate energies follow from diagonalizing the resulting 2×2 Hamiltonian:

$$E_{\pm} = E_0 + \frac{1}{4}\Delta g\mu_B B \pm \sqrt{\left(\frac{1}{4}\Delta g\mu_B B\right)^2 + {\Delta_R}^2} \quad (2)$$

Here $\Delta_R$ is the Rashba-induced dark–bright coupling matrix element; unlike the Zeeman term, it has no simple closed-form expression within this minimal model and is treated as a phenomenological parameter. A fully microscopic k·p treatment of $\Delta_R$ has been developed for achiral 2D perovskite quantum dots,[48] where it is expressed in terms of the Rashba coefficient $\alpha_R$ and confinement-dependent envelope-function overlaps; extending that framework to our chiral 2D geometry, in combination with the chiral spin-texture formalism of Sercel et al.,[22] remains an open theoretical task. When $\Delta_R$ is much larger than $\frac{1}{4}\Delta g\mu_B B$ - as expected given the large chiral splitting in our system - the eigenvalues expand to leading order as $E_0 \pm \Delta_R$ plus a residual term proportional to $(\Delta g\mu_B B)^2/\Delta_R$. Because $\Delta g\mu_B B$ changes sign under field reversal while its square does not, this residual term is even in B: the linear, field-direction-dependent contribution is suppressed, and only a field-magnitude-dependent ($B^2$) term survives. This is the mathematical origin of the field-direction invariance summarized above, and it is structurally analogous to the diamagnetic $B^2$ term in Eq. (1), although the two contributions have distinct physical origins - Rashba-induced dark-state mixing here versus polaron-Rashba effective-mass renormalization there. We emphasize that this two-level model is a minimal construction motivated by the qualitative mechanism of Isarov et al.[9] and the chiral exciton framework of Sercel et al.[22], rather than a result taken directly from either study.

**Rashba spin splitting.** Strong SOC at the Pb 6*p*-derived conduction-band minimum, combined with inversion symmetry breaking, produces Rashba-type spin splitting in the $[PbI_6]^{4-}$ sublattice.[24-28] The most striking and as yet unexplained feature of the field-dependent data is the non-monotonic RC- polarization energy in the *R*-form: a blueshift of +0.8 meV up to ~7.6 T, followed by a monotonic redshift reaching −2.6 meV at 50 T. This behavior is unique to the dominant emission polarization of the *R*-form and has no counterpart in the LC polarization or in either polarization of the *S*-form. The physical origin of this crossover remains an open question. The Rashba–polaron framework of Shin et al.[19] for 3D $MAPbX_3$ provides qualitative context - competing diamagnetic and polaron-renormalization contributions of opposite sign can produce a non-monotonic field response - but a quantitative account for the 2D chiral geometry requires a theoretical treatment that simultaneously incorporates structural chirality, Rashba SOC, and polaron formation in the quantum-well potential. Such a framework does not yet exist. Resolving this feature experimentally will require systematic measurements as a function of temperature, excitation density, and chiral cation identity, to disentangle the contributions of the chiral crystal field from those of the polaron and Rashba terms. [19,36,37,38]

**Polaron renormalization.** The polar $[PbI_6]^{4-}$ lattice supports Fröhlich large polarons that renormalize the carrier effective mass and introduce a characteristic polaron length scale $r_p$[39-43]. As the magnetic length $\ell_B = \sqrt{\hbar/eB}$ decreases with increasing field, the cyclotron radius $r_c$ approaches $r_p$ and the polaron decouples from the lattice. This polaron decoupling crossover is estimated by Shin et al. to occur near ~20 T for $MAPbCl_3$, where the cyclotron radius at 20 T (~4.0 nm) is close to the large polaron radius.[19] This decoupling demodifies the effective mass and can contribute to the high-field energy shift in our 2D system. The monotonic redshift without any initial blueshift in the *S*-form may reflect a smaller initial diamagnetic coefficient in that sample, consistent with a smaller CPL splitting and possibly a different octahedral tilting angle.

In summary, we have performed circularly polarized magneto-photoluminescence measurements on the enantiomeric pair $R$- and $S$-$(C_4H_9NH_3)_2PbI_4$ in pulsed magnetic fields up to 50 T at 2.0 K. The principal experimental findings are: (1) strong CPL anisotropy with opposite helicity for the two enantiomers (RC/LC ≈ 6.0 for $R$-form; LC/RC ≈ 2.4 for $S$-form) that is robust against an applied 50 T field; (2) field-direction invariance of the PL peak energies at 50 T in both enantiomers, attributed to Rashba-induced mixing of the bright exciton with the nearby $m_x = 0$ dark state, which pins the transition energy against field reversal - a mixing driven by the centro-asymmetric Pb-I-Pb bond-angle distortion imposed by the chiral ammonium cations[11] and analogous to the mechanism identified by Isarov et al.[9] in 0D $CsPbBr_3$ nanocrystals - though direct theoretical confirmation for 2D chiral systems remains outstanding; (3) anomalous redshift of all exciton transitions at high fields, with a non-monotonic (initial blueshift to ~7.6 T, then redshift) behavior unique to the dominant RC polarization of the $R$-form that is absent in the $S$-form; and (4) field-induced enhancement of PL intensity and narrowing of linewidth in all polarizations.

These results establish that chiral 2D perovskites exhibit a rich interplay between structural chirality, Rashba spin splitting, and polaron renormalization that produces qualitatively distinct MPL behavior compared to their achiral counterparts. The field-direction invariance of the CPL emission is particularly significant: we propose it can be rationalized as an extension of the dark-state mixing mechanism reported by Isarov et al. for the field-magnitude dependence of $CsPbBr_3$ nanocrystals up to 8 T, applied here to field-direction reversal across the full 0 ~ 50 T range. This opens possibilities for magnetically robust spin-selective optoelectronic applications.[44,45,46,47] The non-monotonic RC- polarization energy in the $R$-form (initial blueshift followed by redshift, with a turnover near ~7.6 T) remains an open question whose resolution will require a theoretical framework that simultaneously accounts for structural chirality, Rashba SOC, and polaron formation in the 2D quantum-well geometry. Future work should systematically vary the chiral cation structure and the halide composition to map the relationship between the degree of octahedral distortion, the CPL splitting, and the high-field energy shift behavior, and to clarify the physical origin of this non-monotonic feature.

## Experimental Methods

Hydriodic acid (57 wt% in distilled $H_2O$), hypophosphorous acid solution (50 wt% in $H_2O$), (R)-(−)-sec-Butylamine (99%), (S)-(+)-sec-Butylamine (99%) were bought from Aldrich. Lead (II) iodide (99.9985%, metals basis) was purchased from ThermoFisher. All chemicals were used without further purification. Single crystals of *R*- and *S*-$(C_4H_9NH_3)_2PbI_4$ were grown by slow cooling from hydriodic acid solution. Stoichiometric quantities of $PbI_2$ and enantiopure *R*-(+)- or *S*-(−)-1-butylamine were dissolved in hydriodic acid with a trace amount of hypophosphorous acid solution at 90 °C and the solution slowly cooled down to room temperature with 2 °C/h rate. The resulting thin platelet crystals were harvested, washed with anhydrous hexane, and dried under vacuum. Enantiomeric purity and phase identity were confirmed by circular dichroism spectroscopy and powder X-ray diffraction, respectively.

MPL measurements were performed in the Faraday geometry (magnetic field **B** applied along the crystal *c*-axis, parallel to the optical axis) using a capacitor-bank-driven pulsed magnet at the Institute for Solid State Physics (ISSP), University of Tokyo, with a peak field of 50 T and a pulse duration of approximately 35 ms. Samples were mounted on a fiber-coupled optical probe immersed in pumped liquid helium at $T = 2.0$ K. PL spectra were acquired using an electron-multiplying CCD (EMCCD) in spectra-kinetics mode with a time resolution of 400 μs per spectrum, allowing time-resolved spectral collection during the field pulse.

Circular polarization analysis was implemented by placing a cryogenic achromatic quarter-wave plate (QWP) and a fixed linear polarizer in the optical path between the sample and the fiber entrance. Right-circularly polarized (RC, $\sigma^+$) and left-circularly polarized (LC, $\sigma^-$) emission components were distinguished by reversing the magnetic field direction ($\boldsymbol{B}(+\hat{z})\uparrow$ versus $\boldsymbol{B}(+\hat{z})\downarrow$) while keeping the optical elements, i.e. circular polarizer fixed, following the method of Shin et al.[19] This avoids spurious polarization artifacts associated with mechanical rotation of optical elements at cryogenic temperatures. In achiral perovskites such as those studied by Shin et al. [19], reversing the field direction switches the spin selection between RC and LC emission components via the Zeeman effect. We adopted the same field-reversal method here with the expectation that it would likewise interchange the two circular polarization components. However, contrary to this expectation, the PL peak positions and intensities in Figure 2 are identical for $\boldsymbol{B}(+\hat{z})\uparrow$ and $\boldsymbol{B}(+\hat{z})\downarrow$ in both enantiomers, within our spectral resolution. This field-direction invariance demonstrates that the CPL anisotropy in these chiral perovskites is an intrinsic property of the crystal chirality rather than a consequence of the Zeeman spin selection, and rules out any artifact of the detection scheme as its origin.

AUTHOR INFORMATION

**Corresponding Author**

Yongmin Kim - Department of Physics, Dankook University, Cheonan 31116, Korea

orcid.org/0000-0001-9274-723X; Email: yongmin@dankook.ac.kr

**Authors**

Dae Young Park - Department of Science Education, Jeonju University, Jeonju 55069, Korea

orcid.org/0000-0002-6162-9553

Sein Park - Department of Physics, Dankook University, Cheonan 31116, Korea

Nojoon Myoung - Department of Physics Education, Chosun University, Gwangju 61452, Korea

Zhuo Yang - Institute for Solid State Physics, The University of Tokyo, Kashiwa 277-8581, Japan

orcid.org/0000-0002-6656-2239

Yoshimitsu Kohama - Institute for Solid State Physics, The University of Tokyo, Kashiwa 277-8581, Japan

Yasuhiro H. Matsuda - Institute for Solid State Physics, The University of Tokyo, Kashiwa 277-8581, Japan orcid.org/0000-0002-7450-0791

**Author Contributions**

Y.K. conceived and supervised this project; S.P. analyzed the data; D.Y.P. grew the chiral perovskite single crystals; N.M. guide the theoretical supports; Z.Y., Y.K. and Y.H.M. carried out the magneto-optical measurements.

**Notes**

The authors declare no competing financial interest.

ACKNOWLEDGMENT

This work was supported by the National Research Foundation of Korea (NRF) grant funded by the Korea government (MSIT) RS-2025-16065850. The work at the ISSP was partially supported by JSPS KAKENHI, Grant-in-Aid for Transformative Research Areas (A) 23A201, 1000-Tesla Science, No.JP23H04860. Z.Y. was supported by research grant from Research Foundation for Opto-Science and Technology, JSPS KAKENHI Grant Number 25K17324 (Grant-in-Aid for Early-Career Scientists), and 26H00591(Grant-in-Aid for Transformative Research Areas (A)) NM is supported by research fund from Chosun University (2024).